\documentclass[prl,twocolumn,showpacs,preprintnumbers,amsmath,amssymb]{revtex4}
\usepackage[dvips]{graphicx}
\usepackage[dvips]{color}
\usepackage{amsmath}
\usepackage{amssymb}
\usepackage{dcolumn}
\usepackage{dcolumn}
\usepackage{bm}

\begin{document}

\title{Shear induced breakup of droplets in a colloidal dispersion}

\author{Hideki Kobayashi$^{1,2}$}
\author{Hiroshi Morita$^{2}$}

\affiliation{$^{1}$Theoretical Soft Matter and Biophysics,
Institute of Complex Systems and Institute for Advanced Simulation,
Forschungszentrum Juelich, 52425 Juelich, Germany}

\affiliation{$^{2}$Soft Matter Modeling Group, Nanosystem Research Institute,
National Institute of Advanced Industrial Science and Technology,
Umesono 1-1-1 Tsukuba 305-8568, Japan}

\date{\today}

\begin{abstract}
 We present numerical results for the breakup of a pair of colloidal
 particles enveloped by a droplet under shear flow. 
 The smoothed profile method is used to accurately account for the
 hydrodynamic interactions between particles due to the host fluid.
 We observe that the critical capillary number, $Ca_{\rm B}$, at which
 droplets breakup depends on a velocity ratio, $E$, defined as the ratio of the
 boundary shift velocity (that restores the droplet shape to a sphere) to 
 the diffusive flux velocity in units of the particle radius $a$.
 For $E < 10$, $Ca_{B}$ is independent of $E$, as is consistent with the
 regime studied by Taylor \cite{Taylor_1, Taylor_2}.
 When $E > 10$, $Ca_{B}$ behaves as $Ca_{\rm B} = 2E^{-1}$, 
 which confirms Karam and Bellinger's hypothesis \cite{Karam}.
 As a consequence, droplet breakup
 will occur when the time scale of droplet deformation $\dot{\gamma}^{-1}$
 is smaller than
 the diffusive time scale $t_{D} \equiv a^{2}/L\tau$ in units of $a$,
 where $L$ is the diffusion
 constant and $\tau$ is the 2nd order coefficient of the Ginzburg-Landau type free energy of the binary mixture.
 We emphasize that the breakup of droplet dispersed particles is
 not only governed by a balance of forces.
 We find that velocity competition is one of the important contributing factor.
\end{abstract}


\maketitle


 A binary fluid system, such as a polymer blend, evolves under shear
 flow principally through the deformation and breakup of its constituent
 droplets \cite{structure_exp}.
 Sheared binary fluids with dispersed colloidal
 particles are a topic of growing interest because they form the basis
 of many industrial products,
 for instance rubber products, plastic articles and some foods
 containing cornstarch or flour.
 
 Breakup of Newtonian droplets suspended in a Newtonian fluid was
 studied and first understood by Taylor \cite{Taylor_1, Taylor_2}.
 He found, theoretically, in these purely viscous Newtonian systems that the deformation
 and breakup of droplets, in the absence of inertial forces, is
 governed by the capillary number
 $Ca=\eta_{\rm c}R\dot{\gamma}/\sigma$
 and the viscosity ratio
 $p = \eta_{\rm d}/\eta_{\rm c}$,
 where $\eta_{\rm c}$ is the viscosity of the continuous phase, R is the
 droplet radius, $\dot{\gamma}$ is the shear rate, $\sigma$ is the interfacial
 tension and $\eta_{\rm d}$ is the viscosity of the droplet.
 The capillary number $Ca=\eta_{\rm c}R\dot{\gamma}/\sigma$
 is the ratio of the viscous stress exerted on a droplet by
 shear flow to the interfacial tension that attempts to restore
 the droplet's shape to a sphere.
 When the viscous stress is larger than the interfacial tension, the
 droplet becomes unstable and breaks.
 The capillary number at which this breakup occurs is known as the
 critical capillary number, $Ca_{\rm B}$. Taylor suggested that $Ca_{\rm B}$ is
 a function of $p$.

 Taylor's result has been confirmed experimentally
 \cite{Karam, Mason_1, Mason_2, droplet_exp1, droplet_exp2} 
 and numerically \cite{droplet_sim1, droplet_sim2}.
 In the experimental works \cite{Karam, Mason_1, Mason_2, droplet_exp1, droplet_exp2},
 the behavior of $Ca_{B}$ is in reasonable quantitative agreement.
 According to these results, $Ca_{B}$ exhibits a minimum in
 the range $0.1<p<1$. For $1<p$, within the highly viscous
 regime, $Ca_{B}$ increases with
 $p$ until $p\approx4$. For $p>4$, droplet cannot be broken by
 shear.
 For $0.01<p<0.1$, where the viscosity $\eta_{\rm d}$ is low,
 $Ca_{B}$ monotonically increases with decreasing $p$.
 In the highly viscous regime, the rigidity of droplets increases with
 increasing $p$ and finally the droplets behave almost like a rigid
 body. Thus, droplets cannot be broken by shear flow.
 In the low viscosity regime, Karam and Bellinger \cite{Karam}
 hypothesized that $Ca_{B}$'s behavior derives from the shape of
 deformed droplets prior to breakup.
 They argued that a deformed droplet could be stabilized by the internally circulating
 fluid within it. 
 Without internal circulation the deformed droplet is unstable and is likely to pinch
 off and divide due to interfacial tension.
 Internal circulation
 counteracts pinching-off by building the pressure against
 dimples, pushing them outwards. This behavior is typical of 
 softmatter systems in which slow and fast dynamics coexist.

 Evolution of the phase separation of binary fluid dispersed colloidal
 particles has been investigated 
 by many researchers
 \cite{neutron_1, neutron_2, wet_sim_1, wet_sim_2, wet_sim_3,
 wet_sim_4, depletion1}.
 Introduction of colloidal
 particles strongly affects the phase structure.
 Neutral colloidal particles prefer to sit at the domain
 interface \cite{neutron_1, neutron_2}. When the volume fraction of
 colloidal particles is sufficiently large, domain growth stops
 when the interfacial region is filled with particles.  As a result, the final phase
 structure is a gel \cite{neutron_2}.
 Colloidal particles introduced into a binary fluid suppress the hydrodynamic flow induced
 by interface motion if particles favor one of the fluid phases.
 As a result, phase separation 
 slows down \cite{wet_sim_1, wet_sim_2, wet_sim_3, wet_sim_4}.
 With high concentrations of colloidal particles, attractive interactions,
 called wetting-induced depletion interactions, 
 act between particles and thus the final structure becomes a network of
 particles connected by one phase \cite{depletion1}.

 Droplets of a colloidal dispersion under shear 
 have been investigated experimentally \cite{shear_wet_exp_1, shear_wet_exp_2}
 and numerically \cite{shear_wet_sim_1, shear_wet_sim_2}.
 Inertia of the particles adsorbed to the droplet interface
 causes the droplets to deform more strongly. As a result,
 droplets breakup more easily and $Ca_{\rm B}$ decreases in comparison
 to a Newtonian droplet \cite{shear_wet_sim_1}. 
 In two dimensional simulations with low particle concentrations,
 the presence of dispersed particles or enveloped polymers
 serves to reduce the deformation of the droplet at intermediate $Ca$. However,
 long polymers induce droplet breakup at high $Ca$ \cite{shear_wet_sim_2}.
 The mechanisms governing the breakup of droplets in the presence of
 dispersed particles are complex and have not been revealed completely
 until now.
 
 In a droplet of a binary fluid dispersed colloid under
 shear, internal circulation is strongly affected by the impermeable
 surface of the colloidal particles. We hypothesize that
 the presence of particles suppresses the diffusive flux
 within the droplet. This effect leads to droplet instability.
 We analyze this effect and discover a new
 mechanism that governs droplet breakup.
 To accomplish this, we have devised a new methodology to directly simulate
 colloidal dispersions in a sheared binary fluid based on the smoothed
 profile method \cite{spm, spm2}.


  We extend the smoothed profile method \cite{spm, spm2} for the
  dynamics of colloidal dispersions in Newtonian fluids to sheared binary fluids.
  In this method, boundaries between solid particles and solvents are
  replaced with a continuous interface by assuming a smoothed
  profile. This enables us to calculate the hydrodynamic interactions both
  efficiently and accurately, without neglecting many-body interactions.
  
  We employ the following free energy functional for a binary mixture
  containing particles:
  
  \begin{eqnarray}
   F\{\psi, \phi\}=\int \bm{{\rm dr}}\{
    f(\psi) - W \psi |\bm{\nabla} \phi|^2 - \chi (\Delta \psi)^2 \phi
    \},
    \label{free_energy}
  \end{eqnarray}
  where $\psi$ is a concentration field and $\phi$ is a smoothed profile
  function. The smoothed profile function $0\leq\phi(\bm{r},t)\leq1$
  distinguishes between the fluid and particle domains, with $\phi=1$
  in the particle domain and $\phi=0$ in the fluid domain. These domains
  are separated by thin interfacial regions with a thickness 
  characterized by $\xi$. 
  The first term on the right-hand side of
  equation (\ref{free_energy}) corresponds to a Ginzburg-Landau type
  mixing free energy for a binary fluid: 
  $f(\psi) = -\tau \psi^2/2 + u \psi^4/4 + K |\bm{\nabla}\psi|^2/2$,
  where $\tau, \mu$ and $K$ are constants. The second term represents the
  wetting interaction between the binary fluid and the particle surface,
  which is represented by $|\bm{\nabla}\psi|^2$. $W$ expresses the strength of
  the wetting interaction. With $W>0(W<0)$, the phase with $\psi>0(\psi<0)$
  favours the particle surface. The third term is introduced so that the
  concentration field inside each particle favors a surface
  phase. $\chi$ is the coupling constant and 
  $\Delta\psi=\psi - W/|W|$ $(W\neq0)$, $\psi$ $(W=0)$. 
 
  The time dependence of the concentration field $\psi$ and the fluid
  velocity $\bm{u}$ are given by
  
  \begin{eqnarray}
   \frac{\partial \psi}{\partial t} = 
    -(\bm{u}\cdot\bm{\nabla})\psi +
    L \bm{\nabla} \cdot \left\{ (\bm{I} - \bm{n} \bm{n} ) \cdot \bm{\nabla}\mu \right\}
  \end{eqnarray}
  \begin{eqnarray}
   \rho_{\rm f}  \left\{ \frac{\partial \bm{u}}{\partial t}+ 
		  (\bm{u}\cdot\bm{\nabla})\bm{u} \right\} =
   -\psi\bm{\nabla}\mu
   -\bm{\nabla}p + \eta \bm{\nabla}^2\bm{u} +
   \rho_{\rm f}\phi\bm{f}_{\rm p} + \bm{f}_{\rm shear}
  \end{eqnarray}
  under the incompressibility condition
  $\bm{\nabla}\cdot\bm{u}=0$, where $\mu$ is
  ${\delta F}/{\delta \psi}$, $\bm{I}$ is the unit tensor, $\bm{n}$ is a
  unit vector field defined by $-\bm{\nabla}\phi/|\bm{\nabla}\phi|$
  and $\rho_{\rm f}$, $\eta$, $L$ and
  $p(\bm{r},t)$ are the density, shear viscosity, diffusion
  constant and pressure field of the solvent,
  respectively. The operator $(\bm{I} - \bm{n} \bm{n})$ directly assigns
  the no-penetration condition at interfacial regions.
  Note that, in this paper, we assume that the two phases have the same viscosity and thus
  the viscosity ratio $p=1$. 
  A body force $\phi\bm{f}_{\rm p}$ is
  introduced to ensure rigidity of the particles and to enforce
  non-slip boundary conditions at the fluid/particle interface.
  Mathematical expressions for $\phi$ and $\phi\bm{f}_{\rm p}$ are
  detailed in previous papers \cite{spm,spm2}. The external force
  $\bm{f}_{\rm shear}$ is introduced to maintain a linear shear with
  shear rate $\dot{\gamma}$. This force is applied using an oblique coordinate
  transformation based on tensor analysis \cite{onuki,oblique}.

  In the present study, we treat colloidal particles as beads of
  radius $a$.
  The motion of the $i$th bead is governed by Newton's
  and Euler's equations of motion with:
  
  \begin{eqnarray}
   M_i\frac{d}{dt}\bm{V}_i=\bm{F}_i^{\rm H}+\bm{F}_i^{\rm P},\;\;\;
    \frac{d}{dt}\bm{R}_i=\bm{V}_i,
  \end{eqnarray}
  
  \begin{eqnarray}
   \bm{I}_i\cdot\frac{d}{dt}\bm{\Omega}_i=\bm{N}_i^{\rm H},
  \end{eqnarray}
  where $\bm{R}_i$, $\bm{V}_i$ and $\bm{\Omega}_i$ are the position,
  translational velocity, and rotational velocity of the beads,
  respectively. $M_i$ and $\bm{I}_i$ are the mass and moment of inertia,
  and $\bm{F}_i^{\rm H}$ and $\bm{N}_i^{\rm H}$ are the hydrodynamic
  force and torque exerted by the solvent on the beads, respectively
  \cite{spm,spm2}. 
  $\bm{F}_i^{\rm P}$ represents the potential force due to direct
  inter-bead interactions, such as Coulombic or Lennard-Jones
  potentials. The truncated Lennard-Jones interaction is expressed in terms of $U_{\rm LJ}$:
  
  \begin{eqnarray}
   U_{\rm LJ}(r_{ij})=\left\{
		       \begin{array}{lll}
			4\epsilon \left\{
				   \left( \dfrac{2a}{r_{ij}}\right)^{12} -
				   \left( \dfrac{2a}{r_{ij}}\right)^{6}\right\}
			+ \epsilon &(r_{ij}<2^{\frac{7}{6}}a), \\
			0 &(r_{ij}>2^{\frac{7}{6}}a),
		       \end{array}
			       \right.
  \end{eqnarray}
  where $r_{ij}=|\bm{R}_i - \bm{R}_j|$. The parameter $\epsilon$
  characterizes the strength of the interactions.
  
  Numerical simulations have been performed in three spatial dimensions with
  periodic boundary conditions. The lattice spacing $\Delta$ is set to
  unit length. The phase interfacial length
  $l=\sqrt{K/\tau}=1$ or $2$ and the interfacial tension
  $\sigma=\sqrt{\tau K}=\tau l$. The unit of time is given by
  $\rho_{\rm f}\Delta^2/\eta$, where $\eta=1$ and $\rho_{\rm f}=1$.
  The system size is $L_x\times L_y\times L_z=64\times64\times32$.
  The range of $\dot{\gamma}$ is
  $2.0\times10^{-3}<\dot{\gamma}<4.0\times10^{-3}$ and
  that of $L$ is
  $8.0\times10^{-2}<L<1.8$.
  The remaining parameters
  are as follows:
  $\tau=u=K/l^{2}=0.01\sim1.0$, $\chi=10 W = 100 l \eta \dot{\gamma}$,
  $a=4$ or $5$, $\xi=2$, $\epsilon=1$,
  $M_i=4\pi a^3/3$ and $h=0.064$, where h is the time increment of a
  single simulation step. 

  The initial state of the binary fluid is set according
  to the following procedure.
  At first, two particles are introduced at $(32 - a, 32, 16)$
  and $(32 + a, 32, 16)$ into a homogenerous concentration of average
  concentration $\bar{\psi}=-0.93$ and $\tau=u=K=0.1$, $W=0.04$,
  $\chi=0.4$, $L=1.6$. We then iterate until steady
  state. Afterwards, we increment by a single time step.
  
  In the following simulations, the time dependence of the concentration
  field $\psi$ and the fluid velocity $\bm{u}$ are discretized using
  a de-aliased Fourier spectral scheme in space and an Euler scheme in
  time.
  To follow bead motions, the position, velocity and angular velocity of
  the beads are integrated by the Adams-Bashforth scheme. 
  
  \begin{figure}[tb]
   \begin{center}

    \includegraphics[width=0.6\hsize]{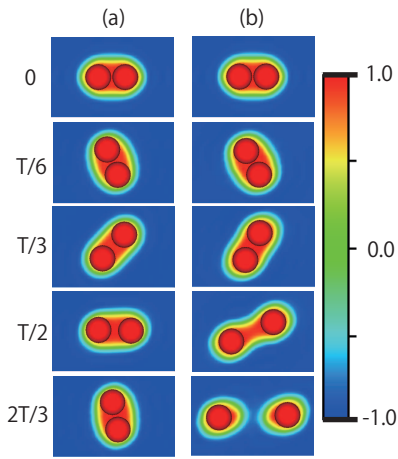}
    \caption{Schematic illustration of the rotation and breakup of a
    droplet.
    $T$ is the tumbling period. The legend relates the color to
    the concentration.
    (a) $Ca$ is sufficiently small and the droplet can rotate. 
    (b) $Ca$ is sufficiently large and
    the droplet breaks up before it can tumble.}
    \label{rotation_breakup}
   \end{center}
  \end{figure}

  Fig. \ref{rotation_breakup} is a schematic illustration of the rotation
  and breakup of a droplet.
  When the interfacial tension is sufficiently large, as in (a), the droplet
  undergoes a tumbling motion similar to a
  chain \cite{jeffrey}, although this tumbling has a limited lifetime.
  On the other hand, in (b), the interfacial tension is small and we expect
  the droplet to break-up before undergoing tumbling.

  We define the breakup shear rate $\dot{\gamma_{\rm B}}$ as the
  shear rate at which a droplet of
  a colloidal dispersion becomes unable to tumble for a given $\sigma$.
  The break up capillary number $Ca_{B}$ is defined as
  \begin{eqnarray}
   Ca_{B}\equiv \frac{2 \eta a \dot{\gamma_{\rm B}}}{\sigma},
  \end{eqnarray}
  assuming that the droplet radius is roughly estimated by $2a$.

  To quantify the barrier to internal circulation due to the particles,
  we introduce the velocity ratio $E$ which is a dimensionless number
  given by the ratio of the
  boundary shift velocity (that restores the droplet shape to a sphere) to 
  the diffusive flux velocity in units of the particle radius $a$.
  Concretely speaking, $E$ is expressed as
  \begin{eqnarray}
    E\equiv\frac{\sigma/\eta}{L\tau/a}=\frac{l a}{\eta L}.
  \end{eqnarray}

  \begin{figure}[tb]
   \begin{center}
    \includegraphics[width=0.6\hsize]{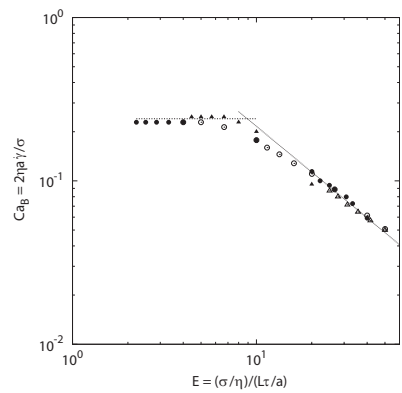}
    \caption{$Ca_{\rm B}$ as a function of $E$.
    Closed circles ($a=4,\dot{\gamma}=2.0\times10^{-3},l=1$). 
    Open circles ($a=4,\dot{\gamma}=4.0\times10^{-3},l=1$).
    Closed triangles ($a=4,\dot{\gamma}=1.0\times10^{-3},l=2$).
    Open triangles ($a=5,\dot{\gamma}=2.0\times10^{-3},l=1$).
    The dashed line corresponds to $Ca_{\rm B}=0.24$. The dotted line
    corresponds to $Ca_{\rm B}=1.84E^{-0.931}$.}
    \label{Ca_vs_E}
   \end{center}
  \end{figure}

  We plot the behavior of $Ca_{B}$ as a function of $E$ in Fig.
  \ref{Ca_vs_E}.
  Were droplet breakup driven purely by the mechanism
  described in earlier works
  \cite{Taylor_1, Taylor_2, Karam, Mason_1, Mason_2,
  droplet_exp1, droplet_exp2, droplet_sim1}, $Ca_{\rm B}$
  would be independent of $E$. For $E < 10$, $Ca_{\rm B}$ is
  constant, namely $Ca_{\rm B}\approx0.24$.
  Previous works \cite{Taylor_1, Taylor_2, Karam, Mason_1, Mason_2,
  droplet_exp1, droplet_exp2, droplet_sim1}
  reported a critical capillary number of $Ca_{\rm B}\approx0.5$ in the
  absence of particles and with equal droplet and solvent viscosities.
  Our value of $Ca_{\rm B}$ is smaller than the previously reported
  result due to the presence of colloidal particles near the interface.
  
  We confirm that the mechanism governing the breakup of droplet
  dispersed colloids in this regime is the same mechanism as described in Taylor's works
  \cite{Taylor_1, Taylor_2}.
  According to Taylor's work \cite{Taylor_1}, a critical
  capillary number $Ca_{\rm B}$ is estimated. We neglect the deviation
  of the droplet shape from a sphere. When the droplet
  radius is roughly approximated by $2a$, the maximum pressure difference across
  the interface is $\delta p \approx 4 \eta \dot{\gamma}$ and the Laplace pressure
  due to interfacial tension is $\delta p_{\rm L} = \sigma / a$.
  The Stokes force exerted on the colloidal particles is estimated as
  $6\pi\eta a d \dot{\gamma}$ where $d$ is the distance between particles, approximated as
  $d=2a$. Thus, the Stokes force per unit area, 
  $\delta F_{\rm S}$, is approximately
  $6\pi\eta a d \dot{\gamma}/ 2 \pi a^{2}=6 \eta \dot{\gamma}$.
  Assuming that the droplet breaks when
  $\delta p + \delta F_{\rm S} > \delta p_{\rm L}$, we can predict $Ca_{\rm B}\approx0.2$.
  This estimate is in good agreement with our result $Ca_{\rm B}\approx0.24$. 
  
  However, when $E > 10$,
  $Ca_{\rm B}$ comes to depend on $E$.
  Although the interfacial tension exceeds the viscous stress,
  the droplet is broken by the shear flow in this
  regime.
  We reveal for the first time that $Ca_{\rm B}$ can be described by
  \begin{eqnarray}
   Ca_{\rm B} = 1.80 E ^{-0.93}.
    \label{ca_b}
  \end{eqnarray}
  This equation implies that $Ca_{\rm B}$ decreases with decreasing
  inner circulation due to the diffusion flux. This behavior
  confirms Karam and Bellinger's hypothesis \cite{Karam}.

  Furthermore,
  Eq. (\ref{ca_b}) not only confirms
  Karam and Bellinger's hypothesis \cite{Karam} but also demonstrates a new
  source of instability originating from the competition between shear flow and
  diffusive flux.
  For droplet dispersed colloidal particles, internal
  circulation, which corresponds to a phase diffusive flux, is obstructed by the particle
  surface.
  The phase diffusive flux takes
  some amount of time, $t_{D} \equiv a^{2}/L\tau$, to overcome this obstruction.
  When $t_{D}$ is larger than the time scale of droplet deformation
  $\dot{\gamma}^{-1}$,
  droplets can break up even when interfacial tension exceeds the
  viscous stress.
  Therefore, we predict that the droplet is unstable when $a^{2}\dot{\gamma}/L\tau > 1$.
  On the basis of this, an instability
  condition can be derived as:
  \begin{eqnarray}
   Ca > 2 E ^{-1},
    \label{theoreticaly_ca_b}
  \end{eqnarray}
  which is in good qualitative agreement with Eq. (\ref{ca_b}).

  \begin{figure}[tb]
   \begin{center}
    \includegraphics[width=0.6\hsize]{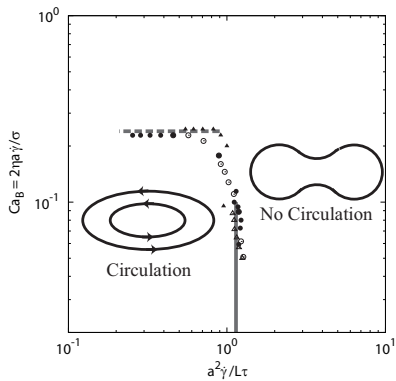}
    \caption{$Ca_{\rm B}$ as a function of $a^2\dot{\gamma}/L\tau$.
    Closed circles ($a=4,\dot{\gamma}=2.0\times10^{-3},l=1$). 
    Open circles ($a=4,\dot{\gamma}=4.0\times10^{-3},l=1$).
    Closed triangles ($a=4,\dot{\gamma}=1.0\times10^{-3},l=2$).
    Open triangles ($a=5,\dot{\gamma}=2.0\times10^{-3},l=1$).
    The dashed line corresponds to $Ca_{\rm B}=0.24$. The solid line
    corresponds to $a^2\dot{\gamma}/L\tau = 1.14.$}
    \label{Ca_vs_lRe}
   \end{center}
  \end{figure}

  To verify the instability condition, 
  $Ca_{\rm B}$ is plotted as a function of $a^{2}\dot{\gamma}/L\tau$ in
  Fig. \ref{Ca_vs_lRe}.
  Fig. \ref{Ca_vs_lRe} shows that the droplet is stable for
  both $Ca < 0.24$ and
  $a^{2}\dot{\gamma}/L\tau < 1.14$.
  Although the interfacial tension exceeds the viscous stress, the droplet is
  unstable and breaks up when $a^{2}\dot{\gamma}/L\tau > 1.14$.
  This result is in good agreement with our prediction, thus validating our
  instability condition.


  The breakup of pairs of colloidal
  particles enveloped by a droplet under shear flow was numerically calculated using the smoothed
  profile method extended to sheared binary fluids for
  $4.0\times10^{-2}<C_{a}<4.0\times10^{-1}$ and
  $2.0\times10^{-1}<a^{2}\dot{\gamma}/L\tau<2.0$.
  Breakup occurs for $0.24 < Ca$ and
  $1.14 < a^{2}\dot{\gamma}/L\tau$.

  Droplet breakup at $Ca \approx 0.24$ occurs because the viscous
  stress exceeds the interfacial tension.

  At $a^{2}\dot{\gamma}/L\tau \approx 1.14$ breakup occurs
  because the time scale of droplet deformation $\dot{\gamma}^{-1}$
  is smaller than $t_{D} \equiv a^{2}/L\tau$.
  We emphasize that this instability is not derived from force competition.
  It is driven by velocity competition.

\end{document}